\documentclass[epj,twocolumn]{webofc}
\usepackage[varg]{txfonts}   % Web of Conferences font
\usepackage[colorlinks=true,allcolors=blue]{hyperref}

\woctitle{FUSION17}
\wocname{EPJ Web of Conferences}
\makeatletter
\def\NAT@def@citea{\def\@citea{\NAT@separator}}
\makeatother

\begin{document}
\title{Time-dependent mean-field investigations of the quasifission process}

\author{\firstname{A.S.} \lastname{Umar}\inst{1}\fnsep\thanks{\email{umar@compsci.cas.vanderbilt.edu}} \and
        \firstname{C.} \lastname{Simenel}\inst{2}\fnsep\thanks{\email{cedric.simenel@anu.edu.au}} \and
        \firstname{S.} \lastname{Ayik}\inst{3}\fnsep\thanks{\email{Ayik@tntech.edu}}
        % etc.
}

\institute{Department of Physics and Astronomy, Vanderbilt University, Nashville, TN 37235, USA
\and
           Department of Nuclear Physics, Research School of Physics and Engineering, The Australian National University, Canberra ACT  2601, Australia
\and
           Physics Department, Tennessee Technological University, Cookeville, TN 38505, USA
          }

\abstract{%
We demonstrate that the microscopic Time-dependent Hartree-Fock (TDHF) theory provides an important approach to
shed light on the nuclear dynamics leading to the formation of superheavy elements.
In particular, we discuss studying quasifission dynamics and calculating ingredients for
compound nucleus formation probability calculations. We also discuss possible extensions
to TDHF to address the distribution of observables.

}
\maketitle
\section{Introduction}
\label{intro}
Synthesis of superheavy elements (SHE)~\citep{dullmann2015} with fusion-evaporation reactions is strongly
hindered by the quasifission (QF) mechanism which prevents the formation of an
equilibrated compound nucleus and depends on the structure of the reactants.
Within the last few years the time-dependent Hartree-Fock (TDHF) approach~\cite{negele1982,simenel2012,nakatsukasa2016} has been
utilized for studying the dynamics of
quasifission~\cite{golabek2009,kedziora2010,simenel2012,simenel2012b,wakhle2014,oberacker2014,hammerton2015,umar2015c,umar2015a,sekizawa2016a,umar2016}
and scission dynamics~\cite{simenel2014a,scamps2015a,simenel2016a,goddard2015,goddard2016,bulgac2016}.
Such calculations are now numerically feasible to perform on a
3D Cartesian grid without any symmetry restrictions
and with much more accurate numerical methods\,\cite{umar2006c,sekizawa2013,maruhn2014,schuetrumpf2016}.

Much progress has been obtained by the community working on developments and applications of the time-dependent mean-field approaches to nuclear dynamics since the last edition of the FUSION conference \cite{scamps2015,sekizawa2015b,simenel2015a,stevenson2015,wakhle2015,washiyama2015b}.
Particularly, the study of quasifission is showing a great promise to provide
insight based on very favorable comparisons with experimental data \cite{wakhle2014,hammerton2015}.
These include comparison with experimental kinetic energy (TKE) and mass-angle distributions (MAD).
In particular, the dependence of quasifission on the orientation of deformed nuclei \cite{hinde1996,hinde2008,nishio2008,nishio2012}, time-scales for
the quasifission process \cite{toke1985,rietz2011,rietz2013}, the connection between the sticking times and the rotation of the compound nucleus \cite{toke1985,thomas2008}, influence of magic shells \cite{itkis2004,knyazheva2007,kozulin2014,wakhle2014}
are some of the experimental insights obtained from the theory.
Similarly, an extension of TDHF called the density-constrained TDHF~\cite{umar2006b,umar2009a}
(DC-TDHF) has
been used to obtain microscopic potential barriers, capture cross-sections, and excitation energies for
superheavy~\cite{oberacker2010,umar2010a} and lighter systems~\cite{umar2006a,umar2012a}.

\section{Insights from TDHF}
\label{sec-2}
\subsection{Capture cross-sections}
\label{subsec-2.1}
The experiments to discover new elements are notoriously difficult, with
fusion evaporation residue (ER) cross-section in pico-barns.
This cross-section is commonly expressed in the product form
\begin{equation}
\sigma _\mathrm{ER}=\sum_{L=0}^{J_\mathrm{\max }}\sigma
_\mathrm{cap}(E_\mathrm{c.m.},L)P_\mathrm{CN}(E^*,L) W_\mathrm{sur}(E^*,L)\;,
\label{eq:er}
\end{equation}
where $\sigma _\mathrm{cap}(E_\mathrm{c.m.},L)$ is the capture cross section at center of mass energy
$E_\mathrm{c.m.}$ and angular momentum $L$. $P_\mathrm{CN}$ is the probability that the composite
system fuses into a compound nucleus (CN) rather than breaking up via quasifission, and
$W_\mathrm{sur}$ is
the survival probability of the fused system against fission.
\begin{figure}[!htb]
	\centering
	\includegraphics*[width=7.5cm]{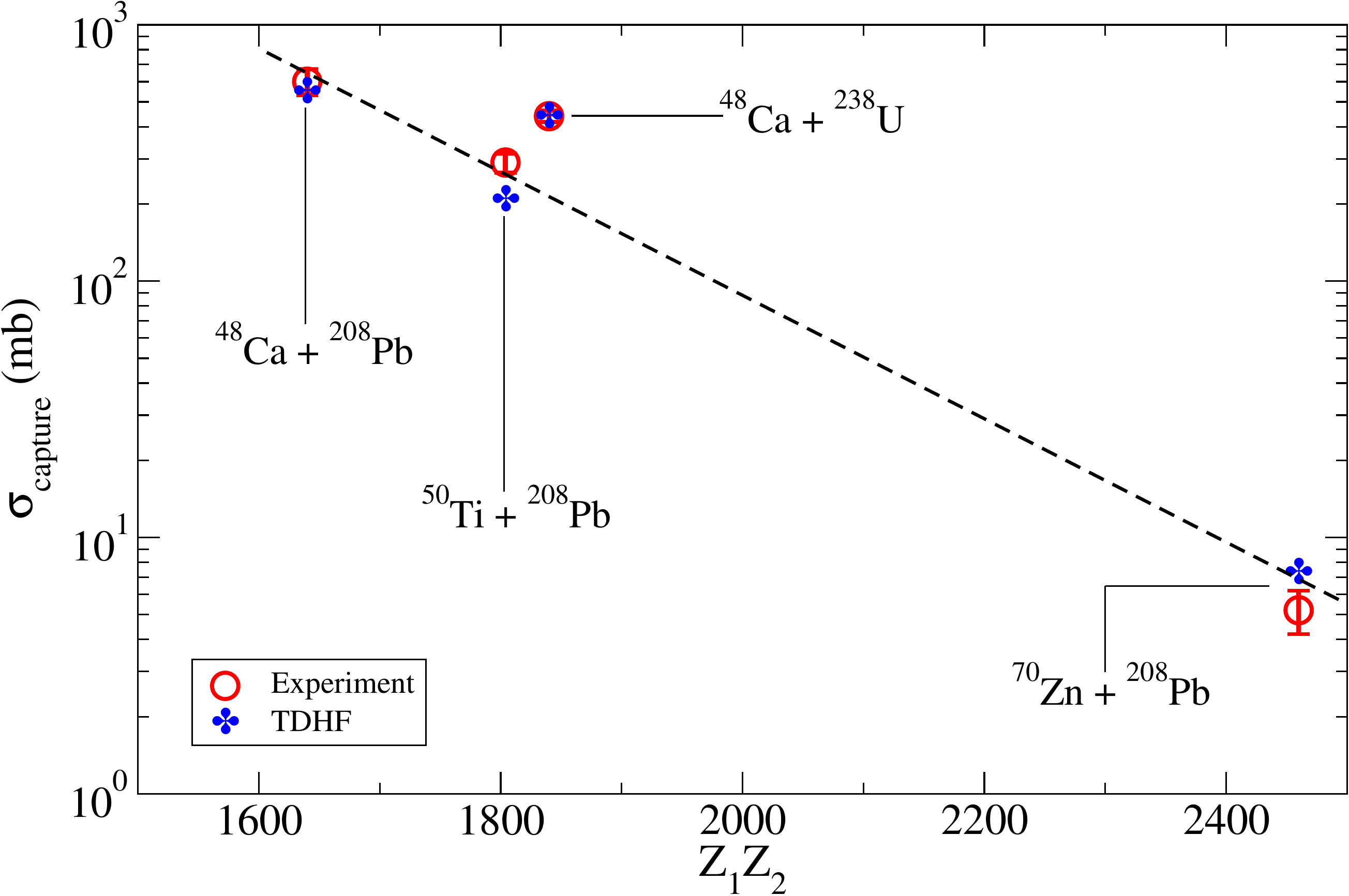}
	\caption{\protect Capture cross-sections calculated using TDHF and the corresponding experimental values plotted
	as a function of $Z_1Z_2$. The dashed line shows a linear dependence for systems with $^{208}$Pb as the target.
	The systems with $^{238}$U target fall on another line with different slope~\cite{kozulin2016,bock1982,toke1985}.}
	\label{fig-0}
\end{figure}

The TDHF time-evolution coupled with the DC-TDHF approach have been used to calculate
capture cross-sections (cross-section for the system to be trapped in a potential pocket)
for a variety of systems~\cite{umar2010a,oberacker2010}. Calculation of capture cross-sections for
reactions involving deformed nuclei are particularly difficult since an average over all
orientations of the deformed nucleus with respect to the beam direction has to be taken~\cite{umar2006a}.
In Fig.\,\ref{fig-0} we plot the calculated capture cross-sections with the corresponding
experimental ones. We see that there is a very reasonable agreement with the experiment
 for these systems.
The dashed line shows a linear dependence for systems with $^{208}$Pb as the target.
The systems with $^{238}$U target fall on another line with different slope~\cite{kozulin2016,bock1982,toke1985}.

\subsection{Mass-angle distributions}
\label{subsec-2.2}
\begin{figure}[!htb]
	\centering
	\includegraphics*[width=8cm]{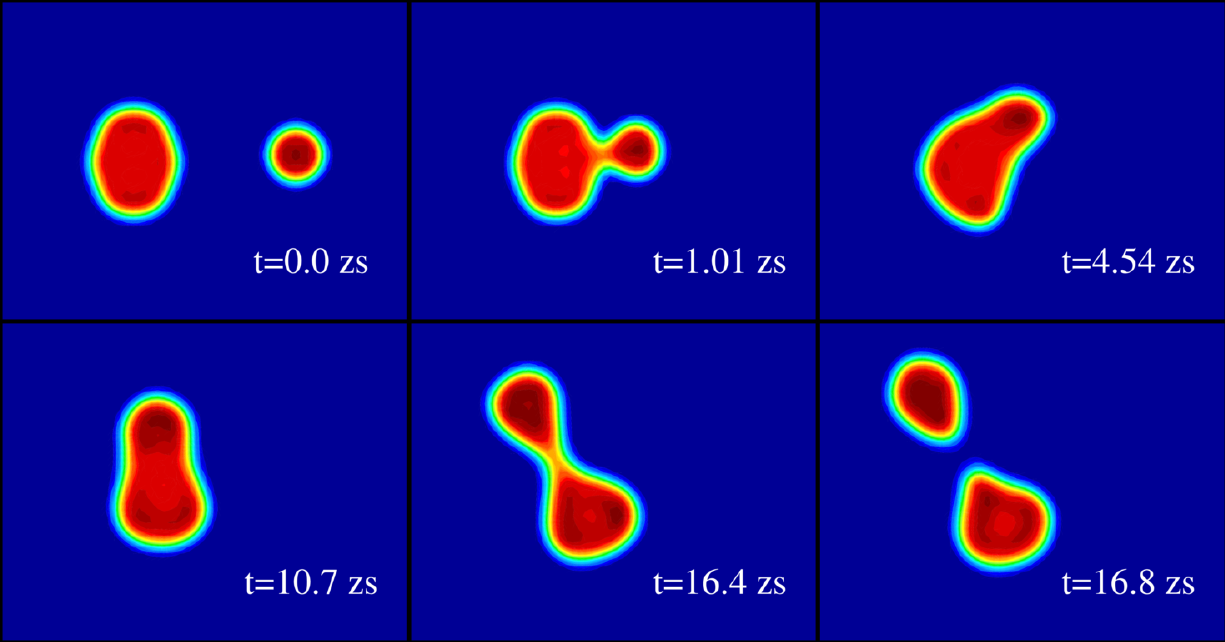}
	\caption{\protect Quasifission in the reaction $^{50}\mathrm{Ti}+^{249}\mathrm{Bk}$
		at $E_{\mathrm{c.m.}}=233$~MeV with impact parameter $b=1.0$~fm.
		Shown is a contour plot of the time evolution of the mass density. Time
		increases from left to right and top to bottom.}
	\label{fig-1}
\end{figure}

Study of quasifission together with capture are intimately related to understanding the
process for forming a compound nucleus, the quantity named $P_\mathrm{CN}$ in Eq.~\ref{eq:er}~\cite{yanez2013}.
Figure~\ref{fig-1} shows the time-evolution of the $^{50}\mathrm{Ti}+^{249}\mathrm{Bk}$ reaction
at $E_{\mathrm{c.m.}}=233$~MeV and impact parameter $b=1.0$~fm~\cite{umar2016}.
For this impact parameter and energy TDHF theory predicts
quasifission.
As the nuclei approach each other, a neck forms between the
two fragments which grows in size as the system begins to rotate.
Due to the Coulomb repulsion and centrifugal forces,
the dinuclear system elongates and forms a very long neck which eventually
ruptures leading to two separated fragments.
In this case, the contact time is found
to be $16$~zs.

\begin{figure}[!htb]
	\centering
	\includegraphics[width=5cm,clip]{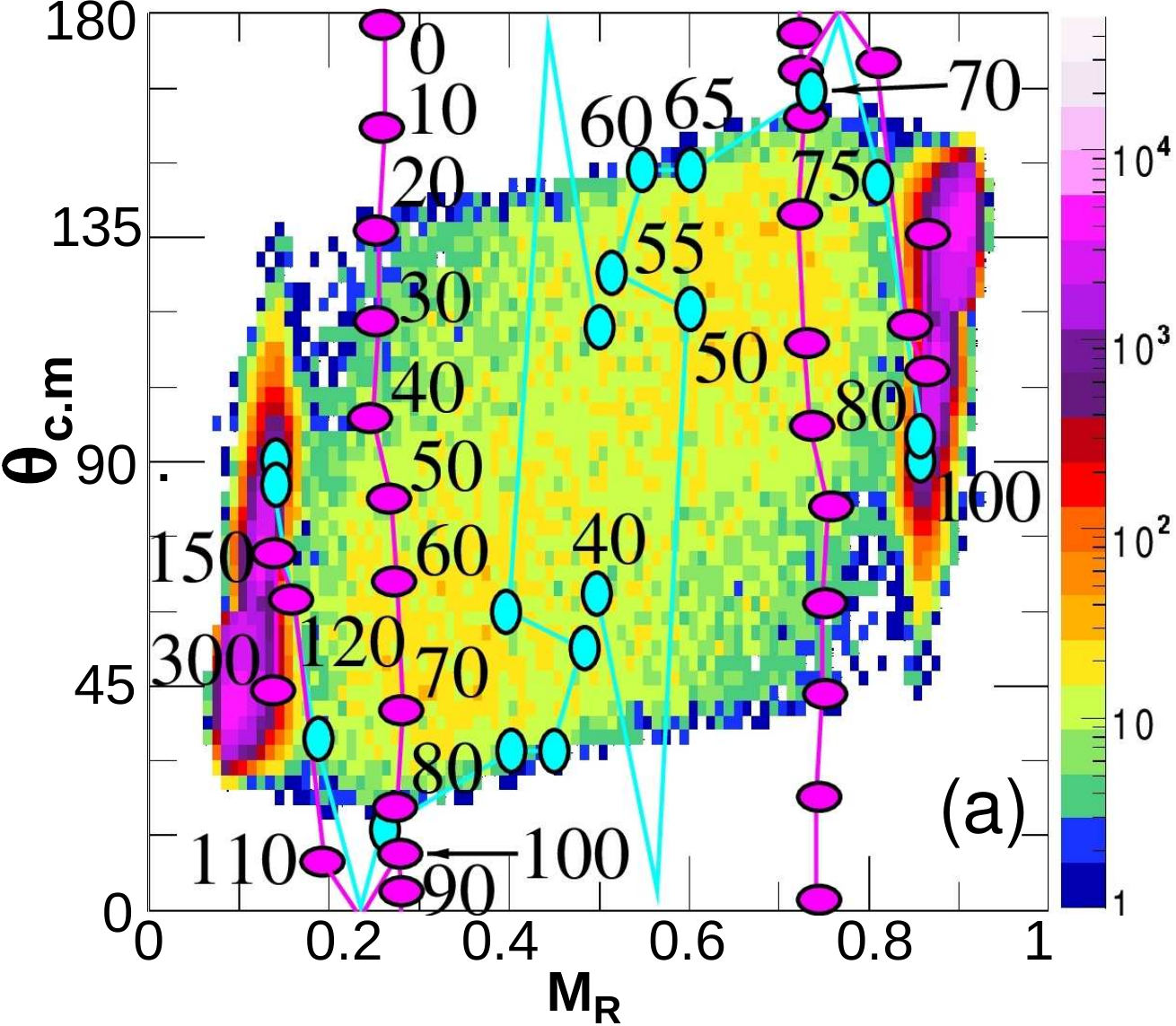}\hspace{0.1cm}\includegraphics[scale=0.34,clip]{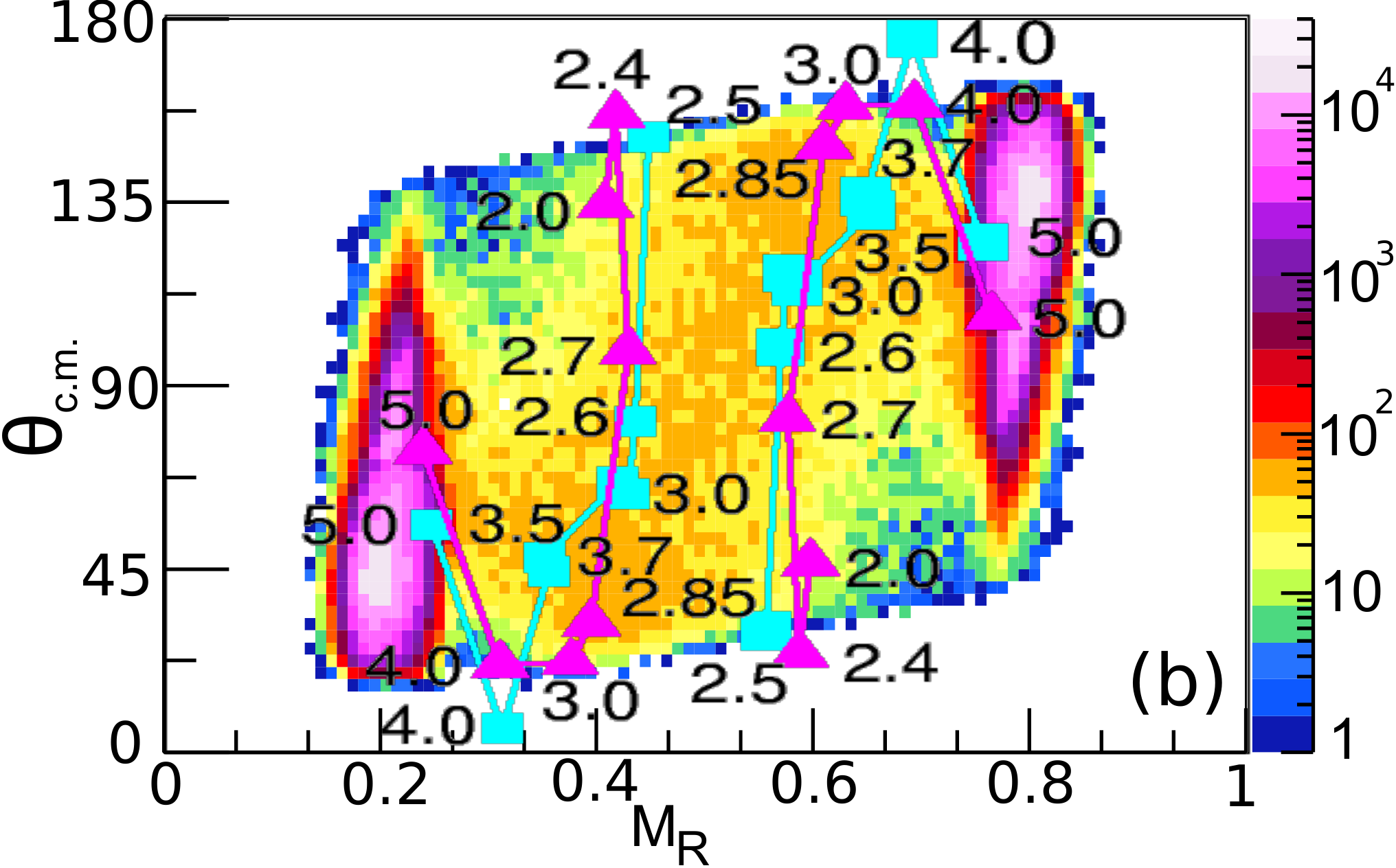}
	\caption{MADs corresponding to two reactions
		(a) $^{40}$Ca+$^{238}$U at energy $E_\mathrm{c.m.}/V_B=1.142$~\cite{wakhle2014}, and (b) $^{54}$Cr+$^{186}$W at
		energy $E_\mathrm{c.m.}/V_B=1.130$~\cite{hammerton2015}, where $V_B$ is the
		corresponding Bass barrier height. The filled symbols
		show the fragments obtained from TDHF calculations marked by the value of the initial orbital angular
		momentum in units of $\hbar$ (a) or the impact parameter in units of fm (b). The density of the experimental
		masses are shown in the adjoint legends.}
	\label{fig-2}
\end{figure}

The utilization of TDHF to study quasifission can serve to gain insights about this reaction
process that is not easily available from the experiments.
Experimentally, the produced fragments in the collisions of relatively heavy systems
originate from deep-inelastic, quasifission, and fusion-fission events. Fragments produced
from deep-inelastic collisions correspond to smaller mass/charge exchange, while the fragments
from the fusion-fission peak around the symmetric breakup of the equilibrated excited compound
nucleus. Fragments coming from the quasifission process fill the region in between but can also
contribute to the symmetric breakup region. Sometimes these are referred to as fast-quasifission (t$\approx$few-35\,zs)
versus slow-quasifission (t>35\,zs). We observe that the quasifission is a dynamical process and does not
correspond to an equilibrated system.
%For two-column wide figures use syntax
One of the best ways to visualize the distribution of produced fragments is through the so-called
mass-angle distribution (MAD) \cite{toke1985,hinde2008}.
In Fig.~\ref{fig-2} we show MADs corresponding to two reactions
(a) $^{40}$Ca+$^{238}$U at energy $E_\mathrm{c.m.}/V_B=1.142$~\cite{wakhle2014}, and (b) $^{54}$Cr+$^{186}$W at
energy $E_\mathrm{c.m.}/V_B=1.130$~\cite{hammerton2015}, where $V_B$ corresponds to the Bass barrier height.
The plot shows the detected fragments with respect to their c.m. detection angle and mass-ratio $M_R=M_1/(M_1+M_2)$.
The filled symbols
show the fragments obtained from TDHF calculations marked by the value of the initial orbital angular
momentum in units of $\hbar$ (a) or the impact parameter in units of fm (b). The multiplicity of the experimental
masses are shown in the adjoint legends.
These reactions illustrate very clearly the deformation dependence of the quasifission products.
For the highly asymmetric reactions involving an actinide nucleus such as the $^{40}$Ca+$^{238}$U system,
the large deformation of Uranium has a major impact on quasifission products. As we see in Fig.\,\ref{fig-2}(a)
the TDHF results originating from collisions with the tip of the Uranium (purple symbols) contribute mainly to the mass asymmetric regions, whereas collisions with the side of the Uranium (cyan symbols) provide significant contribution to the more mass symmetric region.
For the less mass asymmetric and neutron-rich system $^{54}$Cr+$^{186}$W shown in Fig.\,\ref{fig-2}(b) TDHF
contributes less to the mass symmetric region and the effects of deformation has largely disappeared.
The primary reason for the tip orientation resulting in a more asymmetric breakup is due to the fact that
for this orientation nuclei come into contact earlier and rapid excitation builds up while the system
is still in a less compact configuration~\cite{umar2010a}.
We also note that the calculations correctly produce the transition from deep-inelastic to the quasifission regime.
In addition, TDHF calculations also explain the dependence of MADs on the beam energy~\cite{wakhle2014}, as well
as the dependence on the neutron-richness of the target and/or projectile~\cite{hammerton2015}.
\begin{figure}[!htb]
	\centering
	\includegraphics[width=7cm]{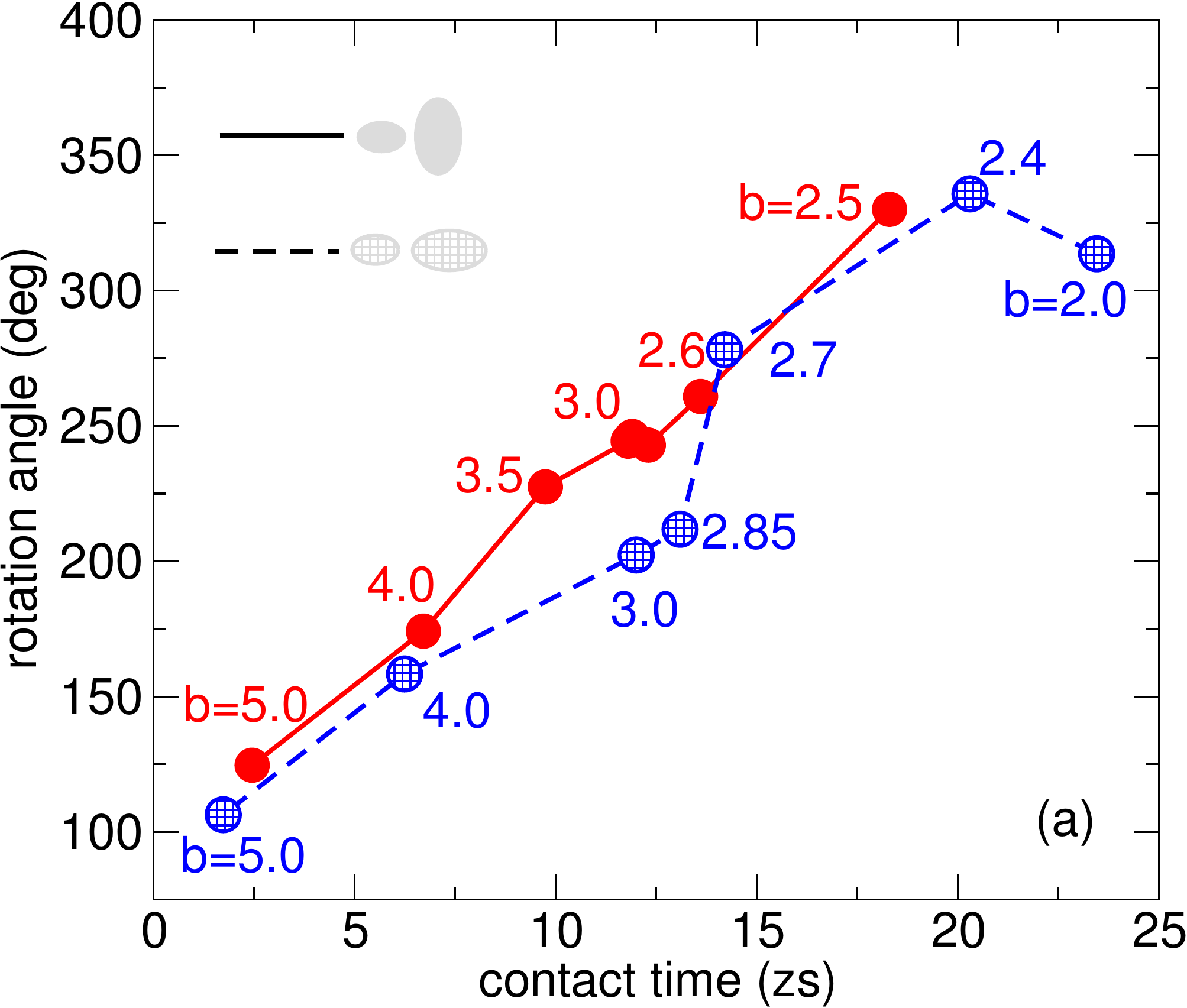}\hspace{0.1cm}\includegraphics[width=7cm]{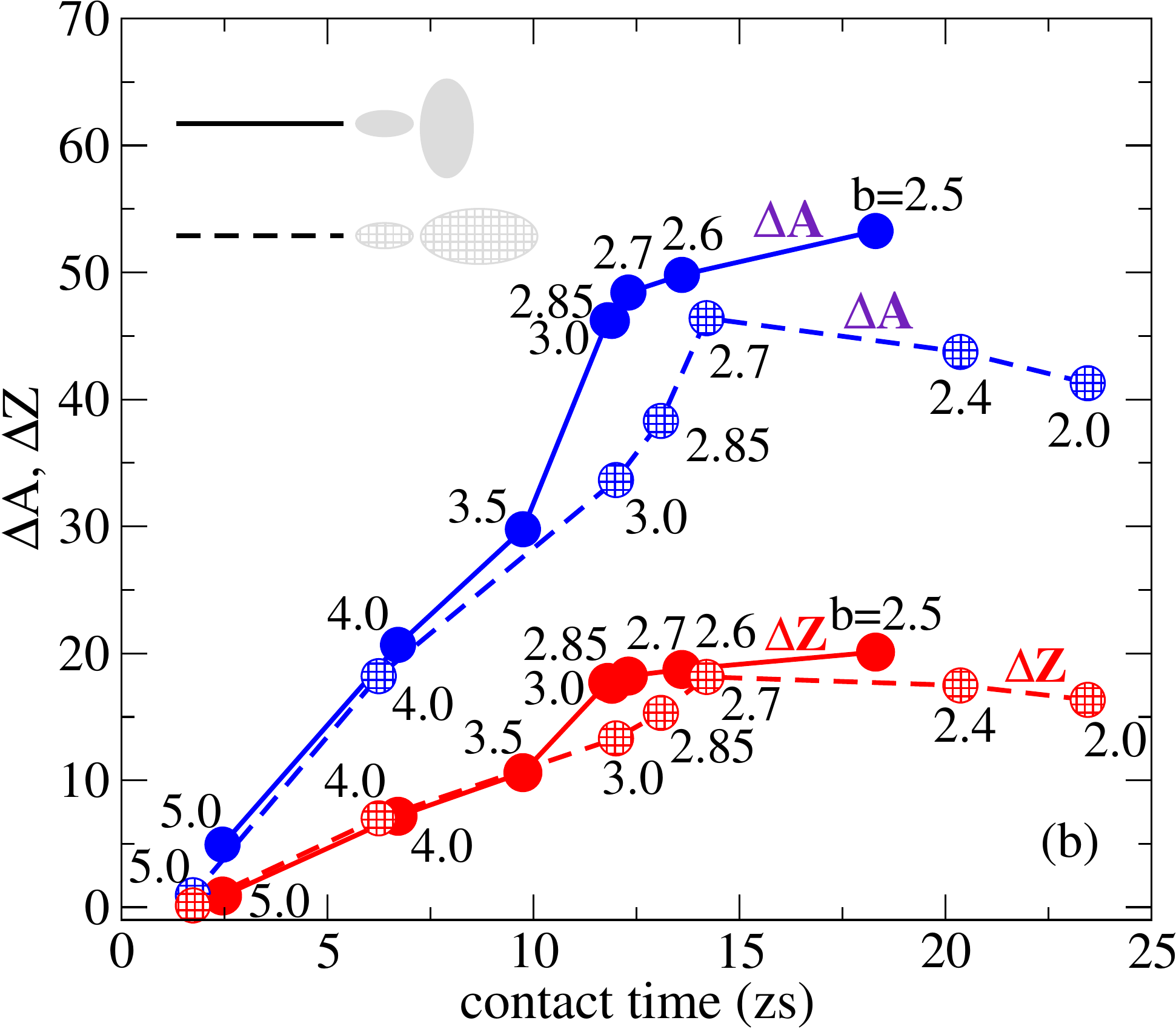}
	\caption{Shown are the
		(a) rotation angle and (b) mass/charge transfer as a function of contact time (zs) for $^{54}$Cr+$^{186}$W at
		energy $E_\mathrm{c.m.}/V_B=1.130$. The numbers adjoint to points indicate the impact parameter in units of fm.}
	\label{fig-3}
\end{figure}

Other quantities that can be studied are the relationship between the mass/charge transfer and rotation angle with
the contact-time (time interval from initial contact to scission). In Fig.~\ref{fig-3} we show these quantities for
$^{54}$Cr+$^{186}$W at energy $E_\mathrm{c.m.}/V_B=1.130$.
In Fig.~\ref{fig-3} we plot the rotation angle (a) and mass/charge transfer (b) as a function of contact time (zs)
for $^{54}$Cr+$^{186}$W at
energy $E_\mathrm{c.m.}/V_B=1.130$. The numbers adjoint to points indicate the impact parameter in units of fm.
We note from Fig.~\ref{fig-3}(a) that the rotation angle generally increases for smaller impact parameters with
the caveat that for small impact parameters the relationship may not be linear i.e. the rotation angle
for $b=2.4$~fm is longer than that for $b=2.0$~fm. This was also observed for other systems~\cite{umar2016}.
A similar behavior is observed in the mass and charge transfer for the tip orientation as seen in Fig.~\ref{fig-3}(b).

\subsection{Fragment TKEs}
\label{subsec-2.3}
\begin{figure}[!htb]
	\centering
    \includegraphics*[width=7cm]{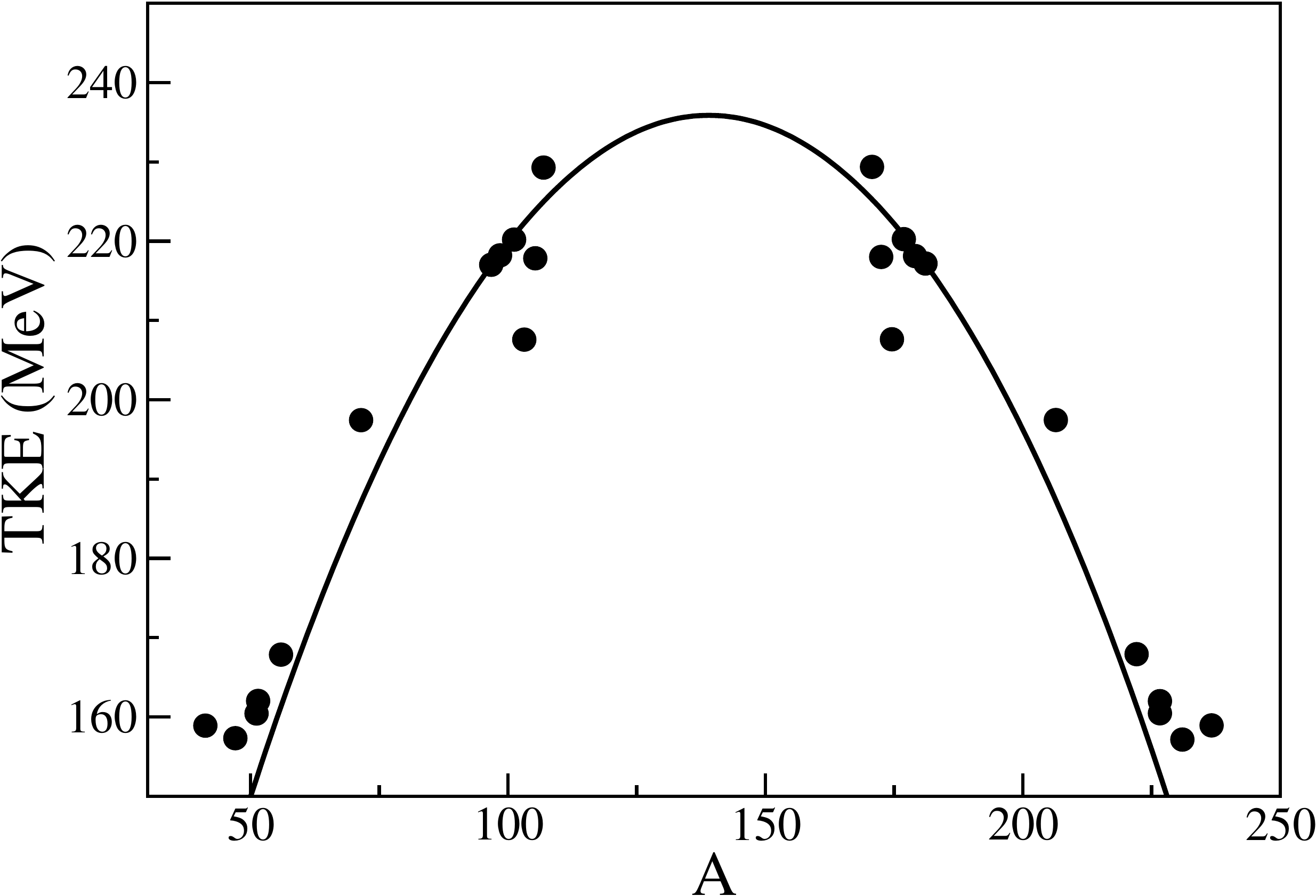}
    \caption{\protect TKE of both the light and heavy fragments formed in $^{40}$Ca+$^{238}$U central collisions at $E_{c.m.}/V_B=1.0-1.1$.
        The solid line represents TKE values based on the Viola formula~\cite{viola1985}.}
    \label{fig:TKE}
\end{figure}

The total kinetic energy of the fission fragments has been investigated experimentally for several systems \cite{itkis2004,knyazheva2007,kozulin2014}
The quasi-fission  contact times are long enough to enable the conversion of the initial relative kinetic energy in to internal excitations. 
Experimentally, the measured total kinetic energy (TKE) of the quasi-fission fragments 
in  $^{40,48}$Ca+$^{238}$U reactions is in relatively good agreement with the Viola systematics~\cite{toke1985,nishio2012}.
The TDHF approach contains one-body dissipation mechanisms which are dominant at near-barrier energy.
It can then be used to predict the final TKE of the fragments. 
The TKE of the fragments formed in $^{40}$Ca+$^{238}$U have been computed for a range of central collisions up to $10\%$ above the barrier.  
Figure~\ref{fig:TKE} shows that the TDHF predictions of TKE are in excellent agreement with the Viola systematics~\cite{viola1985,hinde1992}. 
This indicates that the relative TKE of the quasi-fission fragments
are primarily due to their Coulomb repulsion and do not carry a
fraction of the initial TKE as is the case for deep-inelastic
collisions.

\begin{figure}[!htb]
	\centering
    \includegraphics*[width=7cm]{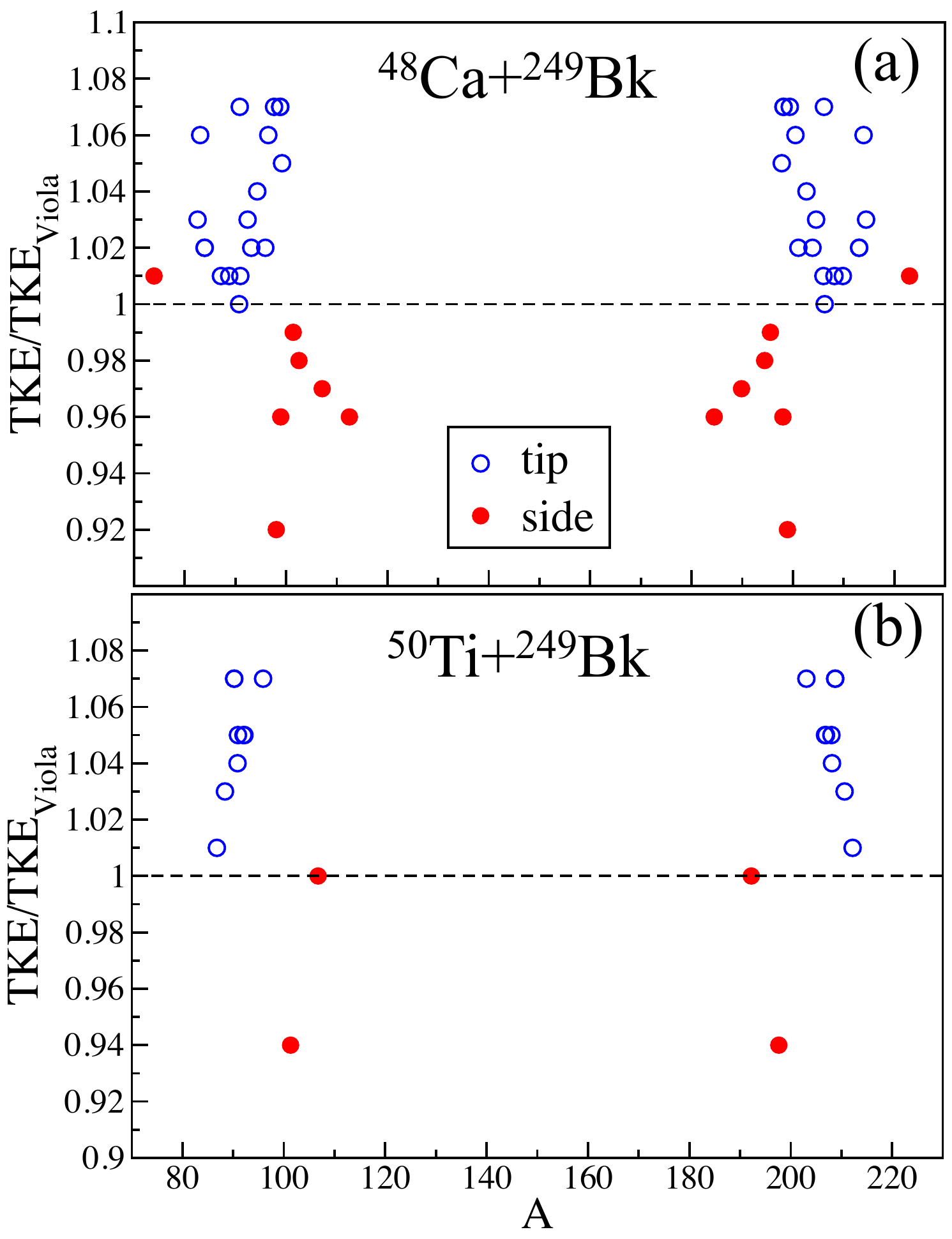}
    \caption{\protect TKE-mass correlations in  (a) $^{48}$Ca+$^{249}$Bk and (b) $^{50}$Ti+$^{249}$Bk. Tip (side) orientations are plotted with open (filled) symbols. The TKE is normalized to the TKE from Viola systematics using the masses and charges of the fragments in the exit channel. Adapted from~\cite{umar2016}.}
    \label{fig:TKE2}
\end{figure}

However, a detailed study of the $^{48}$Ca,$^{50}$Ti+$^{249}$Bk systems also shows an influence of the orientation on the TKE \cite{umar2016}. 
This is illustrated in Fig.~\ref{fig:TKE2} which shows that collisions with the tip are less dissipated than collisions with the side of the actinide.
Recently, very nice results have also been obtained for the $^{64}$Ni+$^{238}$U system~\cite{sekizawa2016} which
compares favorably to the experiment~\cite{kozulin2016}.

\subsection{Shape evolutions}
\label{subsec-2.4}
The proper characterization of fusion-fission and quasifission is one of the most important tasks in analyzing reactions
leading to superheavy elements.
Experimental analysis of fusion-fission and quasifission fragment angular distributions $W(\theta)$  is commonly expressed in terms of
a two-component expression\,\cite{huizenga1969,back1985,back1985a,keller1987,tsang1983},
\begin{equation}
W(\theta) = \sum_{J=0}^{J_\mathrm{CN}}\mathcal{F}_J^{(FF)}(\theta,K_{0}^{2}(\alpha)) + \sum_{J=J_\mathrm{CN}}^{J_\mathrm{max}}\mathcal{F}_J^{(QF)}(\theta,K_{0}^{2}(\alpha)),
\label{eq:wtheta0}
\end{equation}
where
$\alpha\equiv FF$ (fusion-fission) or $QF$ (quasifission).
Here, $J_\mathrm{CN}$ defines the boundary between fusion-fission and quasifission,
assuming a sharp cutoff between the angular momentum distributions of each mechanism.

The quantum number $K$ is known to play an important role in fission\,\cite{vanbenbosch1973}.
The latter is defined as the projection of the total angular momentum along the deformation axis.
In the Transition State Model (TSM)\,\cite{vanbenbosch1973},
the characteristics of the fission fragments are determined by the $K$ distribution at scission.
The argument $K_{0}$ entering Eq.~\ref{eq:wtheta0} is the width of this distribution which is assumed to be Gaussian.
It obeys
$
K_{0}^{2}=T\Im _{eff}/\hbar ^{2}\;,
$
where the effective moment of inertia, $\Im _{eff}$, is computed from the
moments of inertia for rotations around the axis parallel and perpendicular to the %nuclear symmetry axis
principal deformation axis
%\begin{equation}
$\frac{1}{\Im _{eff}}=\frac{1}{\Im _{\parallel }}-\frac{1}{\Im _{\perp }}\;,$
%\label{eq:ieff}
%\end{equation}
and $T$ is the nuclear temperature at the saddle point.
The physical parameters of the fusion-fission part are relatively well known
from the liquid-drop model\,\cite{sierk1986,cohen1974}.
In contrast, the quasifission process never reaches statistical equilibrium.
In principle, it has to be treated dynamically, while Eq.~(\ref{eq:wtheta0}) is based on a statistical approximation.
In addition, the usual choice for the nuclear moment of inertia for
the quasifission component, $\Im _{0}/\Im _{eff}=\text{1.5}$\,\cite{back1985,yanez2013,loveland2015}, is
somewhat arbitrary. Here, $\Im _{0}$ is the moment of
inertia of an equivalent spherical nucleus.

We have developed methods to extract these ingredients directly from TDHF
time-evolution of collisions resulting in quasifission\,\cite{umar2015a}.
The main collective observable of interest for fission and quasifission
(both dynamical and statistical) studies is the moment of inertia of the system.
The proper way to calculate the moment-of-inertia for such
time-dependent densities (particularly for non-zero impact parameters)
is to directly diagonalize the moment-of-inertia
tensor represented by a $3\times 3$ matrix with elements
\begin{equation}
\Im_{ij}(t)/m = \int~d^3r\;\rho(\mathbf{r},t) (r^2\delta_{ij}-x_ix_j)\;,
\end{equation}
where $\rho$ is the local number-density calculated from TDHF evolution, %in units of $(fm^{-3})$,
$m$ is the nucleon mass, and $x_{i=1,2,3}$ denote the Cartesian coordinates.
Numerical diagonalization the matrix $\Im$ gives three eigenvalues.
One eigenvalue corresponds to the moment-of-inertia $\Im_{\parallel}$ for the nuclear system rotating
about the principal axis. The other two eigenvalues define the moments of inertia for
rotations about axes perpendicular to the principal axis.

Using the time-dependent moment-of-inertia obtained from the TDHF collision
one can calculate the so-called effective moment-of-inertia defined above.
It is standard to compute the effective moment of inertia relative to a spherical system
using the mass independent quantity $\Im_0/\Im_{eff}$,
where $\Im_0$ is the
moment-of-inertia of a spherical nucleus with the same number of nucleons\,\cite{tsang1983,yanez2013}.

In our publication\,\cite{umar2015a}
we have calculated the moment-of-inertia ratio for the $^{48}$Ca~+~$^{249}$Bk non-central collisions
at $E_\mathrm{c.m.}=218$~MeV.
At the point of final touching configuration the moment-of-inertia ratios are in the range $1.4-1.8$,
suggesting a relatively strong impact parameter dependence.
We have also studied the energy, impact parameter, and orientation dependence of this ratio and
shown that a simple impact parameter modeling of the ratio may
not be appropriate.
Some of these methods have been utilized in experimental
papers\,\cite{prasad2016}.

\section{Beyond TDHF Calculations}
\label{sec-3}

The mean-field description of reactions using TDHF provides the mean values of the proton and neutron drift.
It is also possible to compute the probability to form a fragment with a given number of nucleons \cite{koonin1977,simenel2010,sekizawa2013,scamps2013a}, but the resulting fragment mass and charge distributions are often underestimated in dissipative collisions \cite{dasso1979,simenel2011}.
Much effort has been done to improve the standard mean-field approximation by incorporating the fluctuation mechanism
into the description. At low energies, the mean-field fluctuations make the dominant contribution to the fluctuation
mechanism of the collective motion.
Various extensions have been developed to study the fluctuations of one-body observables.
These include the TDRPA approach of Balian and V\'en\'eroni~\cite{balian1992}, the time-dependent generator coordinate method~\cite{goutte2005}, or the stochastic mean-field (SMF) method~\cite{ayik2008}.
The effects of two-body dissipation on reactions of heavy systems using the TDDM~\cite{tohyama1985,tohyama2002a}, 
approach have also been recently reported~\cite{assie2009,tohyama2016}.
Here we discuss some recent results using the SMF method~\cite{ayik2015a}.

In the stochastic mean-field (SMF) approach, the fluctuations in the initial
state are incorporated in a stochastic manner by introducing a proper distribution the initial single-particle density
matrices~\cite{ayik2008}. This results in an ensemble of single-particle density matrices generated by evolving each
density in its own mean-field Hamiltonian. %The approach is rather similar to the path integral representation. 
By
calculating the expectation values of an observable in each event, it is possible to determine probability distributions
of observables. In a number of studies, it has been shown that the SMF approach provides a good approximation for
 fluctuations of the collective motion. In particular, in small amplitude limit, the approach gives
rise to the same expression for the dispersion of one-body observables familiar from the variational approach of Balian and V\'en\'eroni~\cite{balian1992}.
\begin{figure}[!htb]
    \includegraphics*[width=7cm]{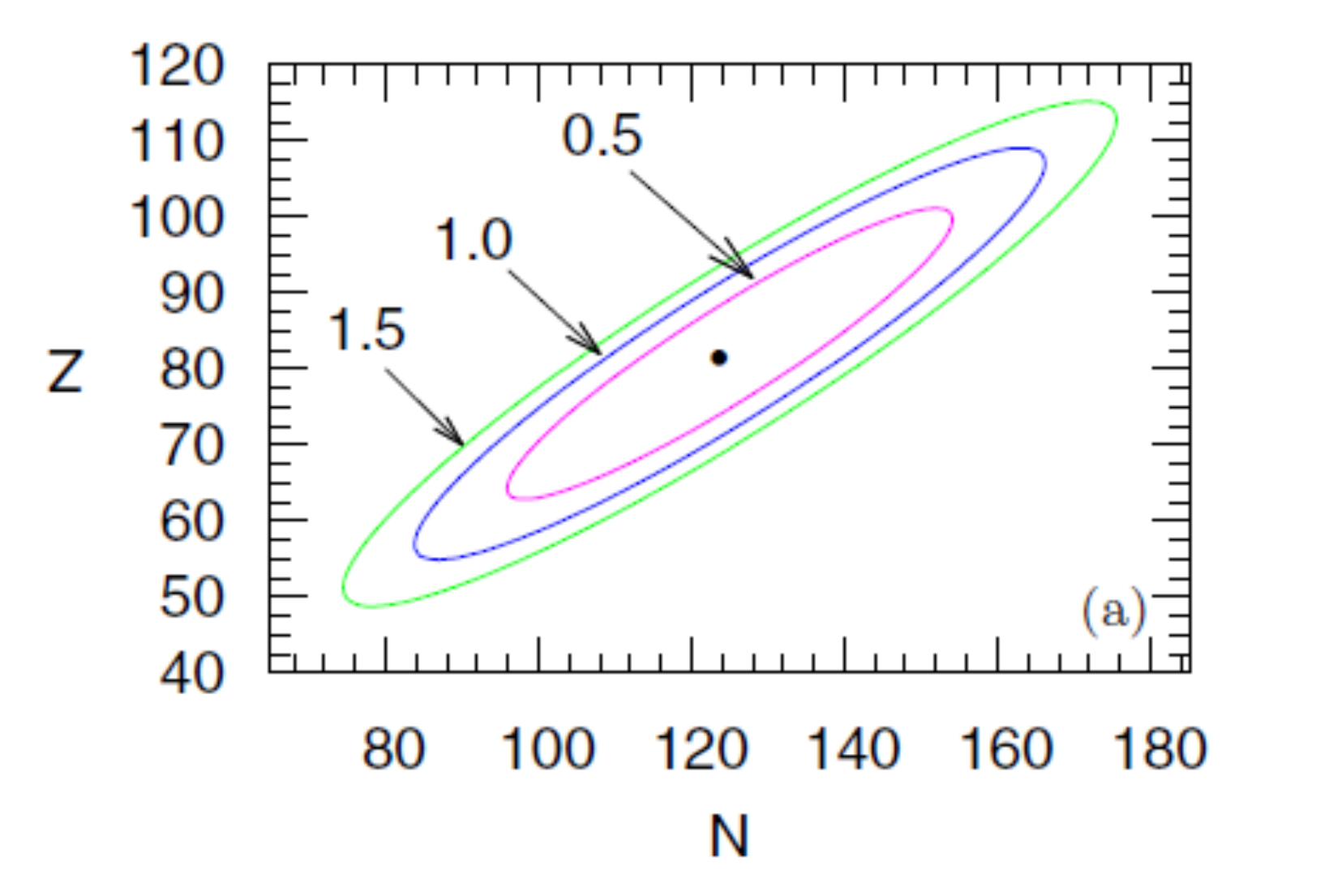}
    \caption{\protect Primary distribution of the target-like fragments produced in central collision of $^{40}$Ca+$^{238}$U
        at $E_{cm}=202$~MeV. Lines indicate equal production probability $exp \left( -C \right)$
        relative to the central fragment with $C=0.5$, $1.0$, and $1.5$.
        Calculated in Stochastic mean-field approach in semi-classical limit.}
    \label{fig:Ayik}
\end{figure}
In the di-nuclear regime, the SMF approach gives rise to Langevin description for nucleon exchange between
projectile-like and target-like nuclei characterized by diffusion and drift coefficients~\cite{ayik2009}. It is well
known that the Langevin description is equivalent to the evolution of the distribution function of the collective
variable according to the Fokker-Plank equation. As a result of this equivalence, instead of carrying out stochastic
simulations, it is more convenient to calculate the transport coefficients and employ the Fokker-Plank approach. When
the drift coefficients are linear functions of mass and charge asymmetry variables, the primary fragment charge and mass
distribution $P \left( N,Z,t \right)$, is given by a correlated Gaussian in the neutron-proton plane. The distribution
function is specified by the mean neutron, $\overline{N}$, and mean proton, $\overline{Z}$, numbers of target-like or
project-like fragments, and co-variances $\sigma_{NN}$, $\sigma_{ZZ}$, and $\sigma_{NZ}$ . The co-variances are
determined by a  set of coupled differential equations in which the inputs are provided by the neutron and proton
diffusion and the drift coefficients.  In the SMF approach these transport coefficients are calculated in terms of the
solutions of the TDHF equations. Calculations take into account the full collision geometry and do not involve any
fitting parameter other than the standard parameters of the Skyrme interaction. In the earlier investigations, transport
coefficients were calculated in the semi-classical approximation in Markovian limit~\cite{yilmaz2014,ayik2015a}.
Recently, we were able to calculate these transport coefficients in the quantal framework by including the shell
structure and the Pauli blocking in an exact manner~\cite{ayik2016}.
As an example, in Fig.~\ref{fig:Ayik} we show the result of nucleon diffusion calculations in the central collisions of $^{40}$Ca+$^{238}$U
system at bombarding energies $E_{cm}=202$~MeV. Figure~\ref{fig:Ayik} illustrates a few equal probability lines for primary population
of the target-like fragments at the exit channel in the $\left( N,Z \right)$
plane. Probability of populating a fragment with neutron and proton numbers $\left( N,Z \right)$
relative to the fragment with mean neutron numbers $\left( \overline{N},\overline{Z} \right)$
is determined by $e^{-C}$, where $C$
indicates numbers on the equal probability lines shown in Fig.~\ref{fig:Ayik}. In this figure dot at the centers of ellipses indicates
the elements with the mean neutron and proton number at the exit channel. The mean values of neutron and proton numbers
at the exit channel are $\left( \overline{N}=125,\overline{Z}=82 \right)$
and $\left( \overline{N}=137,\overline{Z}=86 \right)$ in Ca
induced collisions. For example, the probability of populating a heavy trans-uranium
primary fragment with $  \left( \overline{N}=155,\overline{Z}=98 \right)  $
relative to the probability of populating the element with mean neutron and proton numbers is about $exp \left( -0.5 \right) =0.6$.
We note that the correlated Gaussian function specified by the first two moments provides an approximate description of
the primary fragment population. The approximation is reasonable within the ellipse with $ C=1.0 $
around the center point, but becomes gradually unreasonable as we move out from the center points near to the tail of
the distribution function. For example, as seen beyond the upper ends of the ellipse with $ C=1.5 $,
we observe finite but small probabilities for populating fragments even exceeding the total mass of the system.
Therefore, in particular near the tail region, more accurate description of the fragment population probability is
required.  Also we, note that the primary fragments de-excite by fission and by light particle emission. This
de-excitation process is not incorporated into the calculations.

\section*{Acknowledgements}
This work has been supported by the
Australian Research Council Grants Nos. FT120100760 and DP160101254,
and by the U.S. Department of Energy under grant Nos.
DE-SC0013847 and DE-SC0015513.

%
% BibTeX or Biber users please use (the style is already called in the class, ensure that the "woc.bst" style is in your local directory)
%\bibliography{VU_bibtex_master}
%

\end{document}